\documentclass[11pt]{article}
\usepackage{mathrsfs}

\newcommand{\Eq}[1]{Eq.\ (\ref{#1})}
\newcommand{\Eqs}[2]{Eqs.\ (\ref{#1}) and (\ref{#2})}

\newcommand{\ms}{\mathscr}
\newcommand{\grav}{{\ms G}}
\newcommand{\lslash}[1]{\rlap/#1}

\title{\bf Lorentz-symmetry violating decays in a medium}

\author{\bf Jos\'e F. Nieves\\
Laboratory of Theoretical Physics\\ 
Department of Physics, P.O. Box 23343\\
University of Puerto Rico, R\'{\i}o Piedras,
Puerto Rico 00931-3343
\and
\bf Palash B. Pal\\
Saha Institute of Nuclear Physics\\ 
1/AF Bidhan-Nagar, Calcutta 700064, India
}

\date{December 2007}

\evensidemargin=\oddsidemargin
\makeatletter
\@addtoreset{equation}{section}

\makeatother

\begin{document}

\maketitle

\begin{abstract}
Various decay processes, such as the decay of a spin-1 particle into
two photons or the gravitational decay of a spin-1/2 fermion, are
forbidden in the vacuum by a combination of requirements, including
angular momentum conservation, Lorentz invariance and gauge
invariance. We show that such processes can occur in a medium, such as
a thermal background of particles, even if it is homogeneous and
isotropic.  We carry out a model-independent analysis of the vertex
function for such processes in terms of a set of form factors, and
show that the amplitude can be non-zero while remaining consistent
with the symmetry principles mentioned above.  The results simulate
Lorentz symmetry violating effects, although in this case they arise
from completely Lorentz-invariant physics.
\end{abstract}

\section{Introduction}
It is well known that the rates of physical process that occur in a
medium are modified by the coherent interactions with the background
particles.  It is also now well known that a medium can induce effects
that are not present in the vacuum.  For example, in a medium, a
chiral fermion can obtain an effective mass \cite{Wolfenstein:1977ue,
Weldon:1982bn}, or a Majorana fermion can acquire electric and
magnetic dipole moments \cite{Nieves:1989xg,Semikoz:1989za}, all of
which are forbidden in the vacuum.  Many other similar effects have
been considered in the literature \cite{similareffects}.  In general,
when the particles propagate through a medium, their properties and
interactions are modified such that some processes that are forbidden
in the vacuum can be induced by the effects of the medium.

For our purposes, we can divide such processes in two classes. Some
processes are forbidden in the vacuum for kinematical reasons. That
is, although the off-shell transition matrix element is non-zero, the
process is forbidden for on-shell particles because of energy-momentum
conservation.  However, in the presence of a medium, the dispersion
relations of the particles are modified and those processes can
occur. For example, a free electron cannot radiate a photon in the
vacuum, but in a medium the dispersion relation of the photon makes
the \v{C}erenkov radiation possible.  Another example is provided by
plasmon decay process $\gamma \rightarrow e^+e^-$.  In the vacuum, it
is forbidden due to the masslessness of the photon, but the fact that
the photon dispersion relation is modified in the medium makes the
process possible. We will not be concerned with this class of
processes here.

The other class of processes are those for which the transition matrix
element itself is zero in the vacuum. Invariably, whenever that occurs
it can be attributed to some conservation laws, which in turn are
consequences of the symmetries of the Lagrangian.  It is common to
refer to such processes as being \emph{forbidden}.  However, in
general, a medium is not invariant under the full symmetry group of
the Lagrangian. As a consequence, the corresponding transition
elements can be non-zero when the effects of the background are
included.  Thus, for example, the electric and magnetic dipole moments
of a Majorana fermion can be non-zero in a medium that is
$CPT$-asymmetric~\cite{Nieves:1989xg, Semikoz:1989za}.

A particular subset of the processes in this second class, which are
the focus of this paper, are those which are forbidden in the vacuum
by helicity arguments, or angular momentum conservation.  Consider, for
example, the amplitude for the decay of a spin-0 particle into
another spin-0 particle and a photon.  In the vacuum, the conservation of
angular momentum prevents such processes which, in a more general
form, is the statement that electromagnetic interactions cannot take a
$J=0$ state to another $J=0$ state.  However, the presence of a medium
necessarily breaks the Lorentz symmetry and in particular it breaks
the isotropy of the three-dimensional space that is responsible for
the conservation law that prevents this process from occurring in the
vacuum. As a result, if the transition matrix element is calculated
taking into account the presence of the background, it will not be
zero.

This can be seen in various ways. The medium defines 
a preferred frame in which all
analysis can be performed, namely the frame in which the medium is at rest.
If we commit ourselves to this frame and
carry out the calculations with respect to this frame,
then the expressions for the transition matrix elements are
not restricted by Lorentz invariance, and the terms that are
absent in the vacuum can appear. Alternatively, if the particle
is propagating through the medium, then in the decaying
particle's own rest frame the medium is seen as moving with some
velocity and this, again, breaks the isotropy of the three-dimensional
space.

More generally, we can adopt a completely Lorentz invariant approach by
performing the calculation in an arbitrary frame.  The medium is then
characterized by the temperature and chemical potentials of the
background particles, and by the velocity four-vector of its center of
mass, $v^\mu$.  If the particle is at rest in the medium, then $v^\mu$
is proportional to the particle's momentum vector $p^\mu$ instead of being
an independent vector. Therefore, since the amplitude does not
depend on any additional vectors apart from the momentum or spin or
any other vectors that might characterize the initial and final
states, the obstructions that apply in the vacuum continue to hold
and the amplitude is zero.  However, when the particle propagates
through the medium, the amplitude in general depends on $v^\mu$ in
addition to the various vectors characterizing the initial and final
particles, which invalidates the symmetry argument based on the
isotropy of the three-dimensional space.

In what follows, we adopt the latter point of view. We consider
various decay processes which are forbidden in the vacuum by angular
momentum conservation and/or helicity arguments.
Specifically, we consider the
radiative decay of a spin-0 particle into another spin-0 particle,
the decay of a spin-1 particle into two photons,
the gravitational decay of a spin-0 particle into another spin-0 particle
and the gravitational decay of a spin-1/2 particle into another spin-1/2
particle. In the next sections we review in
each case the arguments that show that the
amplitude for the process in the vacuum vanishes, and then
we demonstrate that the amplitude need not vanish if the process
occurs in a medium. In general,
the presence of the medium also affects the dispersion relations of the
particles appearing in the process.  While those effects may be important
for the calculations of the rates and specific applications,
they are not essential to our arguments and we will ignore them.
Although analogous arguments can be
given for more complicated processes, we have restricted ourselves to
the two-body decay processes mentioned above,
which are straightforward to analyze and for
which the explicit calculations of the transition amplitudes are
simpler. In all our considerations, we assume that the medium is
isotropic and that it can be parametrized in the manner indicated
above. 
The last section contains our conclusions.

\section{Radiative decay of a spin-0 particle into another spin-0 particle}
We first consider a process of the form
\begin{eqnarray}
\phi(p) \rightarrow \phi'(p') + \gamma(q) \,,
\label{process}
\end{eqnarray}
where $\phi$ and $\phi'$ denote scalar (spin 0) particles, $\gamma$
denotes the photon, and $p$, $p^\prime$ and $q$ denote the
corresponding momentum vectors.  

We denote by $j^\mu$ the off-shell electromagnetic vertex, which is
defined such that the on-shell amplitude for the process is given by
\begin{eqnarray}
\ms M = \epsilon^{\ast\mu}(q) j_\mu \,,
\label{sc.Mdef}
\end{eqnarray}
where
\begin{eqnarray}
q = p - p' 
\label{qdef}
\end{eqnarray}
is the photon momentum and $\epsilon^\mu(q)$ is the photon
polarization vector, which satisfies
\begin{eqnarray}
q^\mu \epsilon_\mu(q) = 0\,.
\label{q.e}
\end{eqnarray}
For on-shell particles, $p$, $p^\prime$ and $q$ satisfy the
free-particle dispersion relations,
\begin{eqnarray}
p^2 = M^2 \,, \qquad p'^2 = M'^2 \,,
\label{onshellpp'}
\end{eqnarray}
and
\begin{eqnarray}
q^2 = 0 \,.
\label{qsq}
\end{eqnarray}

For electrically neutral scalar particles, the gauge invariance
condition implies that
\begin{eqnarray}
q^\mu j_\mu = 0 \,,
\label{q.j}
\end{eqnarray}
for any values of $p^\mu$ and $p'^\mu$.  For charged particles, the
condition needs to hold only when $j^\mu$ is evaluated for $p$ and
$p'$ satisfying \Eq{onshellpp'}.  

\subsection{In the vacuum}
In the vacuum, conservation of angular momentum prevents such
processes, as mentioned in the Introduction.  Let us demonstrate how
the result follows, in a manner that will be helpful for analyzing the
corresponding case in a material medium.

We take $p^\mu$ and $q^\mu$ as the independent
momentum variables, using \Eq{qdef} to eliminate $p^\prime$ in favor of them.
Then the vertex function is of the form
\begin{eqnarray}
j^\mu = a_1 p^\mu + a_2 q^\mu\,,
\end{eqnarray}
where the coefficients $a_{1,2}$ are scalar functions of $p$ and $q$.
In this form, \Eq{q.j} implies that
\begin{eqnarray}
\label{vacuumcond}
a_1 p\cdot q + a_2 q^2 = 0\,.
\end{eqnarray}
Since we are considering the vertex function for a transition
amplitude, the fact there is no tree-level electromagnetic coupling
between $\phi$ and $\phi'$ implies that the vertex function is not
singular as $q \rightarrow 0$, and therefore $a$ and $b$ must be of the
form
\begin{eqnarray}
a_1 & = & -q^2 a_3 \,,\nonumber\\
a_2 & = & (p\cdot q) a_3 \,,
\end{eqnarray}
which in turn implies that
\begin{eqnarray}
j_\mu = a_3 \Big[ p\cdot q\, q_\mu - q^2 \, p_\mu\Big]\,,
\end{eqnarray}
with some undetermined scalar function $a_3$.  Thus the on-shell
amplitude, which is calculated with the conditions given in
\Eqs{q.e}{qsq}, is zero.

\subsection{In a medium}
Let us now consider the same process in a background medium.
For our purpose, the crucial difference from the vacuum case is that
in this case the vertex function $j^\mu$ depends also on the velocity
four-vector $v^\mu$ of the medium, and therefore its most general form
is
\begin{eqnarray}
\label{jmediumgen}
j^\mu = a_0 v^\mu + a_1 p^\mu + a_2 q^\mu + a_3 s_\mu \,,
\end{eqnarray}
where
\begin{equation}
\label{sdef}
s_\mu \equiv \epsilon_{\mu\alpha\beta\gamma} p^\alpha q^\beta v^\gamma\,.
\end{equation}
The transversality condition of \Eq{q.j} now implies
\begin{eqnarray}
a_0 q \cdot v + a_1 p\cdot q + a_2 q^2 = 0\,.
\end{eqnarray}
Solving for $a_1$ and substituting the result in \Eq{jmediumgen}
yields
\begin{eqnarray}
j^\mu = a_0\bigg[v^\mu - \frac{q \cdot v}{p\cdot q} \, p^\mu\bigg] +
a_2\bigg[q_\mu - \frac{q^2}{p\cdot q} \,p_\mu\bigg] + a_3 s_\mu\,.
\end{eqnarray}
Using \Eqs{q.e}{qsq} once more, it follows that the term
proportional to $a_2$ does not contribute to the on-shell amplitude
while the $a_{0,3}$ terms yield
\begin{eqnarray}
\ms M = \left({2 a_0\over
M^2 - M'^2}\right) F^\ast_{\mu\nu} v^\mu p^\nu +
a_3 \tilde F^\ast_{\mu\nu} v^\mu p^\nu\,,
\label{Mmedium}
\end{eqnarray}
where $M$ and $M'$ are the masses of the initial and final
scalar particles, respectively.  In writing
\Eq{Mmedium}, we have introduced the notation
\begin{eqnarray}
F_{\mu\nu} = \epsilon_\mu q_\nu - q_\mu\epsilon_\nu 
\end{eqnarray}
and its dual
\begin{equation}
\tilde F_{\mu\nu} = \frac{1}{2}\epsilon_{\mu\nu\alpha\beta}F^{\alpha\beta} \,,
\end{equation}
and used the kinematical relation
\begin{eqnarray}
2p \cdot q = M^2 - M'^2 \,.
\label{p.q}
\end{eqnarray}

Thus, the amplitude is not necessarily zero, and therefore the process
of \Eq{process} can occur in a medium. Written as in \Eq{Mmedium}, it
also reveals why this is so.  The on-shell amplitude should ultimately
contain the factor $F_{\mu\nu}$.  In the case of the vacuum, there is
no antisymmetric tensor, constructed out of the available independent
vectors $p$ and $q$, that gives a non-zero value when contracted with
$F_{\mu\nu}$ to produce a scalar amplitude. In the case of the medium,
there is one such tensor, as \Eq{Mmedium} shows. If the particle is at
rest in the medium, so that $p^\mu$ and $v^\mu$ are parallel, then the
amplitude is zero. This is not unexpected since in that case, in the
rest frame of the particle the three-dimensional space is isotropic
and the angular momentum conservation argument holds again. But if the
particle is moving through the medium, then in its rest frame the
three-dimensional space is not isotropic since there the medium is
seen moving with some velocity, and therefore the angular momentum
conservation argument cannot be applied and the process is not
forbidden.  In addition, the process can occur for charged as well as
electrically neutral scalar particles.

\section{Decay of a spin-1 particle into two photons}
Here consider the process
\begin{eqnarray}
V(k) \to \gamma(q) + \gamma(q') \,,
\label{Zdk}
\end{eqnarray}
where $V$ denotes a massive spin-1 particle. We denote the vertex function
by $\Gamma_{\mu\alpha\alpha'}(q,q')$ in the vacuum case, or
$\Gamma_{\mu\alpha\alpha'}(q,q',v)$ in the presence of the medium,
which is such that the amplitude for the process in any case is given by
\begin{equation}
\label{Vphotonsonshell}
\ms M = \epsilon^{\ast\alpha}(q) \epsilon^{\ast\alpha^\prime}(q^\prime)
\Gamma_{\mu\alpha\alpha'}\varepsilon^\mu(k) \,,
\end{equation}
where $\epsilon^{\alpha}(q)$ is the polarization vector for
a photon with momentum $q$, while $\varepsilon^{\alpha}(k)$
is the corresponding quantity for the $V$ particle. In analogy
with the photon polarization vector, $\varepsilon^\mu(k)$
satisfies
\begin{equation}
\label{varepsilondotk}
k\cdot \varepsilon(k) = 0\,.
\end{equation}

Since all the particles involved in the process are electrically
neutral, electromagnetic gauge invariance implies that
\begin{eqnarray}
q^\alpha \Gamma_{\mu\alpha\alpha'} &=& 0 \,, 
\label{q.Gamma=0} \\*
q'^{\alpha'} \Gamma_{\mu\alpha\alpha'} &=& 0 \,,
\label{q'.Gamma=0}
\end{eqnarray}
for any values of $q$ and $q^\prime$,
in the vacuum as well as in the medium.
In addition the Bose symmetry under the exchange of the two photons
implies that the vertex function satisfies
\begin{eqnarray}
\Gamma_{\mu\alpha\alpha'} (q,q') & = & \Gamma_{\mu\alpha'\alpha} (q',q)\,,
\nonumber\\
\Gamma_{\mu\alpha\alpha'} (q,q',v) & = & \Gamma_{\mu\alpha'\alpha} (q',q,v)\,,
\end{eqnarray}
in the vacuum or in the medium, respectively.
In what follows we apply these conditions to the quantity
\begin{equation}
\label{Vonshellamp}
\Gamma_{\alpha\alpha'}\equiv
\Gamma_{\mu\alpha\alpha'}\varepsilon^\mu(k) \,,
\end{equation}
taking the $V$-boson on shell, as indicated, while maintaining
the photons off-shell.

\subsection{In the vacuum}
As is well-known, the decay represented in \Eq{Zdk} is forbidden in
the vacuum by the combination of angular momentum conservation and the
Bose symmetry between the two photons, a result known as Yang's
theorem \cite{yang}.  Here we review that result in a way that
will help to see how the theorem is evaded in the presence of the medium.
The most general form
of the vertex function allowed by Lorentz invariance
can be read off from an earlier paper\cite{Nieves:1996ff},
\begin{eqnarray}
\label{Zvac.Linv}
\Gamma_{\mu\alpha\alpha'} (q,q') &=& a_1 (q-q')_\mu \eta_{\alpha\alpha'}
+ (a_2 q_\alpha + a'_2 q'_{\alpha}) \eta_{\mu\alpha'}  
+ ( a_3 q_{\alpha'} +
a'_3 q'_{\alpha'}) \eta_{\mu\alpha} \nonumber\\* 
&& + (q-q')_\mu (b_1 q_\alpha q_{\alpha'} + b'_1 q'_\alpha
q'_{\alpha'} + b_2 q'_\alpha q_{\alpha'} + b'_2 q_\alpha q'_{\alpha'})
\nonumber\\* 
&& + c_0 (q-q')_\mu [qq']_{\alpha\alpha'} 
+ (c_1 q_{\alpha} + c'_1 q'_{\alpha}) [qq']_{\mu\alpha'}
+ (c_2 q_{\alpha'} + c'_2 q'_{\alpha'}) [qq']_{\mu\alpha}\,,
\nonumber\\* 
\end{eqnarray}
where $\eta_{\mu\nu}$ is the Minkowski metric tensor, and we have used the
shorthand notation
\begin{eqnarray}
[qq']_{\alpha\alpha'} = \varepsilon_{\alpha\alpha'\beta\gamma} q^\beta
q'^\gamma \,.
\end{eqnarray}
In writing \Eq{Zvac.Linv}, and in what follows, it should be
understood that we are considering the quantity defined in
\Eq{Vonshellamp}, and therefore we omit any term that does not
contribute to that quantity.  Thus, we have avoided the combination
$(q+q')_\mu$ because it vanishes when it is contracted with
$\varepsilon^\mu(k)$.  Some other possibilities, like
$\varepsilon_{\mu\alpha\alpha'\beta} q^\beta$ and
$\varepsilon_{\mu\alpha\alpha'\beta} q'^\beta$, have been omitted
since they are not independent~\cite{Nieves:1996ff}, as can be seen
by contracting the identity
\begin{eqnarray}
g_{\lambda\rho} \varepsilon_{\alpha\beta\gamma\delta} - 
g_{\lambda\alpha} \varepsilon_{\rho\beta\gamma\delta} - 
g_{\lambda\beta} \varepsilon_{\rho\alpha\gamma\delta} - 
g_{\lambda\gamma} \varepsilon_{\rho\alpha\beta\delta} - 
g_{\lambda\delta} \varepsilon_{\rho\alpha\beta\gamma} = 0 
\label{5tensor}
\end{eqnarray}
with various combinations of $q$ and $q'$.

We now apply the transversality conditions stated in
\Eqs{q.Gamma=0}{q'.Gamma=0}.  First of all, they imply
\begin{eqnarray}
a_2 = a'_2 = a_3 = a'_3 = 0 \,,
\end{eqnarray}
and for the other form factors they yield the following relations:
\begin{eqnarray}
a_1 + b_1 q^2 + b_2 q \cdot q' &=& 0 \,, \nonumber\\*
a_1 + b'_1 q'^2 + b_2 q \cdot q' &=& 0 \,, \nonumber\\*
b_1 q \cdot q' + b'_2 q'^2 &=& 0 \,, \nonumber\\* 
b'_1 q \cdot q' + b'_2 q^2 &=& 0 \,, \nonumber\\*
c_1 q^2 + c'_1 q \cdot q' &=& 0 \,, \nonumber\\*
c_2 q \cdot q' + c'_2 q'^2 &=& 0 \,.
\end{eqnarray}
These can be solved, without introducing singularities, by writing
\begin{eqnarray}
a_1 &=& B_1 q^2 q'^2 - b_2 q \cdot q' \,, \nonumber\\*
b_1 &=& - B_1 q'^2 \,, \nonumber\\*
b'_1 &=& - B_1 q^2 \,, \nonumber\\*
b'_2 &=& B_1 q \cdot q' \,, \nonumber\\*
c_1 &=& C_1 q \cdot q' \,, \nonumber\\*
c'_1 &=& -C_1 q^2 \,, \nonumber\\*
c_2 &=& - C_2 q'^2 \,, \nonumber\\*
c'_2 &=& C_2 q \cdot q' \,,
\end{eqnarray}
and substituting these back in \Eq{Zvac.Linv} then yields
\begin{eqnarray}
\Gamma_{\mu\alpha\alpha'} (q,q') &=& (q-q')_\mu \bigg[ B_1
  g^{\beta\gamma} (q^2 \eta_{\alpha\beta} - q_\alpha q_\beta) 
  (q'^2 \eta_{\alpha'\gamma} - q'_{\alpha'} q'_\gamma) \nonumber\\*
&& \hspace{2cm} + b_2 \Big( q'_\alpha q_{\alpha'} - q \cdot q'
  \eta_{\alpha\alpha'}\Big) 
+ c_0 [qq']_{\alpha\alpha'} \bigg] \nonumber\\*
&& + C_1 (q \cdot q' q_\alpha - q^2
  q'_\alpha) [qq']_{\mu\alpha'} + C_2 (q \cdot q' q'_{\alpha'} - q'^2
  q_{\alpha'}) [qq']_{\mu\alpha} \,.
\end{eqnarray}

Bose symmetry implies a relationship between $C_1$ and $C_2$. However,
the terms with the coefficients $B_1$, $C_1$ and $C_2$ do not
contribute to the on-shell amplitude given in \Eq{Vphotonsonshell},
and we need not consider them further.  Regarding the other terms,
recall that the form factors are functions only of the scalar
invariants $q^2$ and $q'^2$, since the other invariant, $q \cdot q'$,
is not independent due to the on-shell condition for the
$V$-boson. The Bose symmetry condition, which requires that $b_2$ and
$c_0$ are odd under the interchange of the two photon momenta, is then
\begin{eqnarray}
\label{Zvac.b2}
b_2 (q^2, q'^2) = - b_2 (q'^2, q^2) \,,
\end{eqnarray}
and similarly for $c_0$. In particular, this implies
that $b_2(0,0) = c_0(0,0) = 0$ for on-shell photons, proving Yang's theorem.

\subsection{In a medium}
In this case, the presence of the vector $v^\mu$ allows us to enumerate
all the possible terms in a more compact way. This is accomplished
by introducing the following vectors,
\begin{eqnarray}
n_\lambda &=& \varepsilon_{\lambda\rho\beta\gamma} v^\rho q^\beta
q'^\gamma \,, \nonumber \\*
t_\lambda &=& \varepsilon_{\lambda\rho\beta\gamma} n^\rho q^\beta
q'^\gamma \,,
\label{nt}
\end{eqnarray}
complemented by the following combinations of $q$ and $q^\prime$,
\begin{eqnarray}
r_\mu & \equiv  & q^2 q^\prime_\mu - (q\cdot q^\prime)q_\mu \,,\nonumber\\
r'_\mu & \equiv & q'^2 q_\mu - (q\cdot q^\prime)q^\prime_\mu \,.
\label{ss'}
\end{eqnarray}
Notice that the vector $n_\mu$, which is analogous to the vector $s_\mu$
defined in \Eq{sdef}, as well as $t_\mu$ are orthogonal to both
$q_\mu$ and $q'_\mu$, while $r_\mu$ and $r'_\mu$ are such that
\begin{eqnarray}
r\cdot q = r' \cdot q' = 0\,.
\label{r.q}
\end{eqnarray}
In addition, the set of vectors
\begin{equation}
A^{(a)}_\mu = \left\{r_\mu, r^\prime_\mu, n_\mu, t_\mu\right\} \qquad
(a = 1,2,3,4)\,,
\end{equation}
are linearly independent in general, and they span the four-dimensional
Minkowski space.  Therefore, like any rank-3 tensor,
$\Gamma_{\mu\alpha\alpha'}$ can be expressed in the form
\begin{equation}
\Gamma_{\mu\alpha\alpha'}(q,q',v) = \sum_{abc} T_{abc}
A^{(a)}_\mu A^{(b)}_\alpha A^{(c)}_{\alpha^\prime} \,,
\end{equation}
but in this case the transversality conditions imply that certain terms
in this expression for $\Gamma_{\mu\alpha\alpha'}$ are actually absent.
The electromagnetic transversality conditions given in
\Eqs{q.Gamma=0} {q'.Gamma=0} imply that $A^{(b)}_\alpha$ and
$A^{(c)}_{\alpha^\prime}$ can take values only from the subsets
$\{r_\alpha, n_\alpha, t_\alpha\}$ and
$\{r^\prime_{\alpha^\prime}, n_{\alpha^\prime}, t_{\alpha^\prime}\}$,
respectively. Moreover, if we consider the $V$ to be on-shell
(i.e., \Eq{Vonshellamp}), then the combination
$(q+q')_\mu$ does not contribute and we need to keep
only one combination of $r_\mu$ and $r^\prime_\mu$ which we take
as $(q - q^\prime)_\mu$. We express all this by writing
\begin{eqnarray}
\varepsilon^\mu(k)
\Gamma_{\mu\alpha\alpha'} (q,q',v) = \varepsilon^\mu(k)\sum_{abc}
T_{abc} \{(q-q')_\mu,t_\mu,n_\mu\}^{(a)}
\{r_\alpha,t_\alpha,n_\alpha \}^{(b)}
\{r'_{\alpha'},t_{\alpha'},n_{\alpha'}\}^{(c)} \,,
\end{eqnarray}
which is the most general form for the vertex function with the $V$
on-shell. For on-shell photons the terms with $r_\alpha$ and/or
$r^\prime_{\alpha^\prime}$ give no contribution, and the physical
amplitude is 
\begin{eqnarray}
\varepsilon^\mu(k)
\epsilon^{\ast\alpha}(q) \epsilon^{\ast\alpha^\prime}(q')
\Gamma_{\mu\alpha\alpha'} (q,q',v) & = &
\varepsilon^\mu(k)
\epsilon^{\ast\alpha}(q) \epsilon^{\ast\alpha^\prime}(q')\nonumber\\
&&\times
\sum_{abc} \hat T_{abc}
\{(q-q')_\mu ,t_\mu ,n_\mu\}^{(a)}
\{t_\alpha,n_\alpha\}^{(b)}
\{t_{\alpha'},n_{\alpha'}\}^{(c)}\,.\nonumber\\
\end{eqnarray}
This contains 12 form factors, and while
Bose symmetry implies certain relations between them,
unlike the case in the vacuum the form factors need not vanish as
a consequence of this.
The reason can be understood by looking at $\hat T_{111}$
as an example. Bose symmetry implies that it is antisymmetric under
the interchange of the two momenta.  However, instead of \Eq{Zvac.b2},
in the present case that condition is
\begin{eqnarray}
\hat T_{111} (q^2,q'^2, q \cdot v,q' \cdot v) = - \hat T_{111}
(q'^2,q^2, q' \cdot v,q\cdot v) \,,
\end{eqnarray}
where we have indicated explicitly the dependence on the various
independent scalar variables. For on-shell photons this yields the relation
\begin{equation}
\hat T_{111} (0,0, q \cdot v,q' \cdot v) = - \hat T_{111} (0,0, q'
\cdot v,q\cdot v) \,,
\end{equation}
which due to the additional dependence on the variables $q \cdot v$ and
$q'\cdot v$ does not imply that the on-shell form factor vanishes.  In general,
therefore, the amplitude for the two-photon decay in a medium does not
vanish.

\section{Gravitational decay of a spin-0 particle into another spin-0
particle} 
In this section, we consider the process
\begin{eqnarray}
\label{process2}
\phi(p) \rightarrow \phi'(p') + \grav(q) \,,
\end{eqnarray}
where $\phi$ and $\phi'$ denote scalars as before, and $\grav$ denotes
a graviton.  We denote by $t^{\mu\nu}$ the off-shell gravitational
vertex, which is symmetric and defined such that the on-shell
amplitude for the process is given by
\begin{eqnarray}
\ms M = \epsilon^{\ast\mu\nu}(q) t_{\mu\nu} \,,
\end{eqnarray}
where $\epsilon^{\mu\nu}$ is the graviton polarization tensor, which
satisfies the relations 
\begin{eqnarray}
\label{gravpolrels}
\epsilon_{\mu\nu} & = & \epsilon_{\nu\mu} \,, \\
q^\nu\epsilon_{\mu\nu}(q) & = & 0 \,, \label{q.egrav}\\
\eta^{\mu\nu}\epsilon_{\mu\nu} & = & 0\,,
\label{egravtrace}
\end{eqnarray}
where $\eta_{\mu\nu}$ is the Minkowski metric as before.

Gravitational gauge invariance implies the transversality conditions
\begin{eqnarray}
\label{q.t}
q^\mu t_{\mu\nu} = q^\mu t_{\nu\mu} = 0 \,,
\end{eqnarray}
but in contrast to \Eq{q.j}, this is required to hold only for
on-shell values of $p$ and $p'$. 

\subsection{In the vacuum}
In the vacuum, we can consider $t_{\mu\nu}$ as a function $p^\mu$ and
$q^\mu$ and write it in the form
\begin{eqnarray}
t_{\mu\nu} = a_0 \eta_{\mu\nu} + a_1 p_{\mu}p_{\nu} + a_2 q_\mu q_\nu +
a_3 \{pq\}_{\mu\nu} \,,
\end{eqnarray}
where the coefficients $a_i$ are scalar functions of $p$ and $q$ and
\begin{eqnarray}
\{p_1 p_2\}_{\mu\nu} \equiv p_{1\mu} p_{2\nu} + p_{2\mu} p_{1\nu}\,,
\end{eqnarray}
for any two vectors $p_{1,2}$. In this form, \Eq{q.t} then implies the
relations
\begin{eqnarray}
a_0 + a_2 q^2 + a_3 p\cdot q & = & 0 \,,\nonumber\\
a_1 p\cdot q + a_3 q^2 & = & 0\,.
\end{eqnarray}
Taking into account the fact there is no tree-level gravitational coupling
between $\phi$ and $\phi'$, which implies that the vertex function is not
singular as $q \rightarrow 0$, we solve these relations by setting
\begin{eqnarray}
a_0 & = & -a_2 q^2 - a_3 p\cdot q \,,\nonumber\\
a_1 & \equiv & -a_4 q^2 \,,\nonumber\\
a_3 & = & a_4 p\cdot q\,,
\end{eqnarray}
which give
\begin{eqnarray}
t_{\mu\nu} = a_2 \left(q_\mu q_\nu - q^2\eta_{\mu\nu}\right) +
a_4\left((p\cdot q)\{pq\}_{\mu\nu} - q^2 p_\mu p_\nu -
(p\cdot q)^2 \eta_{\mu\nu}\right)\,.
\end{eqnarray}
The parameters $a_{2,4}$ remain undetermined, but once again, the
on-shell graviton amplitude, which is calculated with the conditions
given in \Eqs{qsq}{q.egrav}, is zero.

\subsection{In a medium}
In analogy with the photon case the vertex function in this case depends
also on $v^\mu$, and we now write
\begin{eqnarray}
t_{\mu\nu}(p,q,v) & = & a_0 \eta_{\mu\nu} + a_1 p_\mu p_\nu + a_2
q_\mu q_\nu 
+ a_3 \{pq\}_{\mu\nu} + a_4 v_\mu v_\nu + a_5 \{pv\}_{\mu\nu}
+ a_6 \{qv\}_{\mu\nu}\nonumber\\* 
&&\mbox{} + b_1 \{ps\}_{\mu\nu} + b_2 \{vs\}_{\mu\nu}
+ b_3 \{qs\}_{\mu\nu}\,,
\end{eqnarray}
where $s^\mu$ has been defined in \Eq{sdef}.
The transversality condition yields the following relations,
\begin{eqnarray}
a_0 + q^2 a_2 + (p\cdot q)a_3 + (q\cdot v)a_6 & = & 0\,,\nonumber\\
(p\cdot q)a_1 + q^2 a_3 + (q\cdot v)a_5 & = & 0\,,\nonumber\\
(q\cdot v)a_4 + (p\cdot q)a_5 + q^2 a_6 & = & 0\,,\nonumber\\
(p\cdot q)b_1 + (q \cdot v)b_2 + q^2 b_3 & = & 0\,,
\end{eqnarray}
which we use to eliminate $a_{0,1,4}$ and $b_1$ in favor of the others,
and in this way we arrive at
\begin{eqnarray}
t_{\mu\nu}(p,q,v) & = & a_2 \left(q_\mu q_\nu - q^2\eta_{\mu\nu}\right)
+ a_3\left[\{pq\}_{\mu\nu} - (p\cdot q)\eta_{\mu\nu} -
\left(\frac{q^2}{p\cdot q}\right)p_\mu p_\nu\right]\nonumber\\*
&&\mbox{} + a_5\left[\{pv\}_{\mu\nu}
- \left(\frac{p\cdot q}{q \cdot v}\right)v_\mu v_\nu
- \left(\frac{q \cdot v}{p\cdot q}\right)p_\mu p_\nu\right]\nonumber\\*
&&\mbox{} + a_6\left[\{qv\}_{\mu\nu} - (q\cdot v)\eta_{\mu\nu}
- \left(\frac{q^2}{q\cdot v}\right)v_\mu v_\nu\right]\nonumber\\*
&&\mbox{} + b_2\left[\{vs\}_{\mu\nu} -
\left(\frac{q \cdot v}{p\cdot q}\right)\{ps\}_{\mu\nu}\right]
+ b_3\left[\{qs\}_{\mu\nu} -
\left(\frac{q^2}{p\cdot q}\right)\{ps\}_{\mu\nu}\right] \,.
\end{eqnarray}
For an on-shell graviton, the terms proportional to $a_{2,3,6}$ and
$b_3$ vanish, but the terms with the coefficient $a_5$ and $b_2$ give
a non-vanishing contribution to the amplitude,
\begin{eqnarray}
\ms M & = & a_5\epsilon^{\ast\mu\nu}(q)
\left[ \{pv\}_{\mu\nu}
- \left(\frac{p\cdot q}{q \cdot v}\right)v_\mu v_\nu
- \left(\frac{q \cdot v}{p\cdot q}\right)p_\mu
p_\nu\right]\nonumber\\* 
&&\mbox{} + 
b_2 \epsilon^{\ast\mu\nu}(q)\left[\{vs\}_{\mu\nu} -
\left(\frac{q \cdot v}{p\cdot q}\right)\{ps\}_{\mu\nu}\right]\,.
\end{eqnarray}
%

\section{Gravitational decay of a spin-1/2 particle
into another spin-1/2 particle}
We denote by $\Gamma_{\mu\nu}(p,q)$ the off-shell gravitational vertex
function, which is defined such that the amplitude for the process
\begin{eqnarray}
f(p) \rightarrow f^\prime(p^\prime) + \grav(q) \,,
\end{eqnarray}
is given by
\begin{eqnarray}
\ms{M} = \epsilon^{\ast\mu\nu} \bar u_{f^\prime}(p^\prime)
\Gamma_{\mu\nu} u_f(p) \,.
\end{eqnarray}
The vertex function $\Gamma_{\mu\nu}$ is symmetric in its indices, and
gravitational gauge invariance implies that it satisfies the
transversality conditions
\begin{eqnarray}
q^\mu \bar u_{f^\prime}(p^\prime)\Gamma_{\mu\nu}u_f(p) =
q^\mu \bar u_{f^\prime}(p^\prime)\Gamma_{\nu\mu}u_f(p) = 0 \,. 
\end{eqnarray}
In order to write down the general form of the matrix element $\bar
u_{f^\prime}(p^\prime)\Gamma_{\mu\nu}u_f(p)$ we introduce its tensor
and pseudotensor components by writing
\begin{eqnarray}
\bar u_{f^\prime}(p^\prime) \Gamma_{\mu\nu}u_f(p) = 
\bar u_{f^\prime}(p^\prime)\left[\Gamma^{(T)}_{\mu\nu} +
\Gamma^{(P)}_{\mu\nu}\gamma_5\right]u_f(p)\,.
\end{eqnarray}
The transversality condition then implies that
\begin{eqnarray}
q^\nu\bar u_{f^\prime}(p^\prime) \Gamma^{(i)}_{\mu\nu} u_f(p) = 0\,,
\end{eqnarray}
for $i = T,P$ independently. 

\subsection{In the vacuum}
With the understanding that the expression $\Gamma_{\mu\nu}^{(T)} +
\Gamma_{\mu\nu}^{(P)}\gamma_5$ is sandwiched between the spinors, the
most general form of each of the components $\Gamma_{\mu\nu}^{(T,P)}$
is
\begin{eqnarray}
\Gamma^{(i)}_{\mu\nu} = a^{(i)}_0\eta_{\mu\nu} +
a^{(i)}_1 p_\mu p_\nu +
a^{(i)}_2 q_\mu q_\nu +
a^{(i)}_3 \{pq\}_{\mu\nu} +
a^{(i)}_4 \{p\gamma\}_{\mu\nu} +
a^{(i)}_5 \{q\gamma\}_{\mu\nu}\,,
\end{eqnarray}
where we have used the Dirac equation for the spinors to reduce the terms
that contain factors of $\lslash{p}$ or $\lslash{q}$. In addition,
the terms involving
$\epsilon_{\mu\alpha\beta\gamma}p^\alpha q^\beta \gamma^\gamma$
can be reduced to the ones above by using the identity
\begin{eqnarray}
\label{3gammaident}
\gamma_\alpha\gamma_\beta\gamma_\gamma = 
(\eta_{\mu\alpha} \eta_{\beta\gamma} - \eta_{\mu\beta}
\eta_{\alpha\gamma} + \eta_{\mu\gamma} \eta_{\alpha\beta}) \gamma^\mu
+ i\epsilon_{\alpha\beta\gamma\mu}\gamma^\mu\gamma_5 \,,
\end{eqnarray}
together with the Dirac equation.

Let us consider the tensor ($T$)
component first.  Using the Dirac equation for the spinors to reduce
the factors of $\lslash{q}$ that appear, the transversality conditions
imply
\begin{eqnarray}
a_0^{(T)} + q^2 a^{(T)}_2 + (p\cdot q) a^{(T)}_3 +
(m - m^\prime)a^{(T)}_5 & = & 0\,,\nonumber\\
(p\cdot q) a^{(T)}_1 + q^2 a^{(T)}_3 +
(m - m^\prime) a^{(T)}_4 & = & 0\,,\nonumber\\
(p\cdot q) a^{(T)}_4 + q^2 a^{(T)}_5 & = & 0\,,
\end{eqnarray}
To satisfy the last relation we set
\begin{eqnarray}
a^{(T)}_4 & = & -q^2 a^{(T)}_6 \,,\nonumber\\*
a^{(T)}_5 & = & (p\cdot q) a^{(T)}_6 \,,
\end{eqnarray}
and then we use the first and second relations to
solve for $a^{(T)}_0$ and $a^{(T)}_1$, respectively, obtaining
\begin{eqnarray}
\label{a1a2fermioncase}
a^{(T)}_0 & = & -q^2 a^{(T)}_2 - (p\cdot q) a^{(T)}_3 -
(m - m^\prime)(p\cdot q) a^{(T)}_6 \nonumber\\
a^{(T)}_1 & = & \frac{q^2}{p\cdot q}\left((m - m^\prime)a^{(T)}_6 -
a^{(T)}_3\right)\,.
\end{eqnarray}
For the ($P$) component the results are similar, the only difference
being that in the formulas analogous to \Eq{a1a2fermioncase}, instead
of $m - m^\prime$ the factor $-m - m^\prime$ appears.  Thus, for
either component,
\begin{eqnarray}
\Gamma^{(i)}_{\mu\nu} & = &
a^{(i)}_2 \left(q_\mu q_\nu - q^2\eta_{\mu\nu}\right)
\mbox{} +
a^{(i)}_3\left[\{pq\}_{\mu\nu} - \frac{q^2}{p\cdot q}p_\mu p_\nu -
(p\cdot q)\eta_{\mu\nu}\right]\nonumber\\*
&&\mbox{} +
a^{(i)}_6\left\{
q^2\left[\frac{\lslash{q}}{p\cdot q}p_\mu p_\nu -
  \{p\gamma\}_{\mu\nu}\right] 
+ (p\cdot q)\left[\{q\gamma\}_{\mu\nu} -
\lslash{q}\eta_{\mu\nu}\right]\right\}\,,
\end{eqnarray}
where we have used the Dirac equation once again to rewrite the
factors of $m \pm m^\prime$ in terms $\lslash{q}$.  The relations in
\Eqs{q.egrav}{egravtrace} then imply that, for an on-shell graviton,
the amplitude is zero.

\subsection{In a medium}
In this case, the terms involving the factors $\lslash{v}$,
$\sigma_{\alpha\beta} v^\beta$ and 
$\epsilon_{\mu\alpha\beta\gamma}\ell^\alpha v^\beta \gamma^\gamma$
(where $\ell = p,q$), must be taken into account. Therefore we write,
for $i = T,P$ as before,
\begin{eqnarray}
\Gamma^{(i)}_{\mu\nu} & = & 
\eta_{\mu\nu}(a^{(i)}_0 + b^{(i)}_0\lslash{v}) +
p_\mu p_\nu(a^{(i)}_1 + b^{(i)}_1\lslash{v}) +
q_\mu q_\nu(a^{(i)}_2 + b^{(i)}_2\lslash{v}) +
v_\mu v_\nu(a^{(i)}_3 + b^{(i)}_3\lslash{v})\nonumber\\*
&&\mbox{} +
\{pq\}_{\mu\nu}(a^{(i)}_4 + b^{(i)}_4\lslash{v}) +
\{pv\}_{\mu\nu}(a^{(i)}_5 + b^{(i)}_5\lslash{v}) +
\{qv\}_{\mu\nu}(a^{(i)}_6 + b^{(i)}_6\lslash{v})\nonumber\\*
&&\mbox{} +
\{p\gamma\}_{\mu\nu}(a^{(i)}_7 + b^{(i)}_7\lslash{v}) +
\{q\gamma\}_{\mu\nu}(a^{(i)}_8 + b^{(i)}_8\lslash{v}) +
\{v\gamma\}_{\mu\nu}(a^{(i)}_9 + b^{(i)}_9\lslash{v})\,,
\end{eqnarray}
where we have chosen to write the terms with two $\gamma$ matrices in
terms of $\gamma_\mu\lslash{v}$ rather than $\sigma_{\alpha\beta}
v^\beta$.  In addition, the terms with the epsilon symbol mentioned
above have been omitted since they reduce to those that are included
here by using the identity \Eq{3gammaident} and the Dirac equation.

As in the vacuum case, the transversality condition must be satisfied for
$\Gamma^{(T,P)}_{\mu\nu}$ separately, and furthermore, within each
group, the terms with and without $\lslash{v}$ must vanish
separately as well. In this way we obtain the following relations,
\begin{eqnarray}
a^{(i)}_0 & = & -q^2 a^{(i)}_2 - p\cdot q a^{(i)}_4 -
q\cdot v a^{(i)}_6 - \lslash{q} a^{(i)}_8 \,,\nonumber\\
a^{(i)}_1 & = & \frac{-1}{p\cdot q}\left[
q^2 a^{(i)}_4 + q\cdot v a^{(i)}_5 - \frac{\lslash{q}}{p\cdot q}
\left(q^2 a^{(i)}_8 + q\cdot v a^{(i)}_9\right)\right] \,,\nonumber\\
a^{(i)}_3 & = & \frac{-1}{v\cdot q}\left[
p\cdot q a^{(i)}_5 + q^2 a^{(i)}_6 + \lslash{q}a^{(i)}_9\right]\,,
\nonumber\\
a^{(i)}_7 & = & \frac{-1}{p\cdot q}\left[q^2 a^{(i)}_8 + q\cdot v a^{(i)}_9
\right] \,,
\end{eqnarray}
with analogous relations for the set of coefficients $b^{(i)}_j$,
and we obtain the final form
\begin{eqnarray}
\label{Gammagravmedium}
\Gamma^{(i)}_{\mu\nu} & = &
(-q^2\eta_{\mu\nu} + q_\mu q_\nu)(a^{(i)}_2 +
b^{(i)}_2\lslash{v})\nonumber\\* 
&&\mbox{} + \left[
-p\cdot q\eta_{\mu\nu} - \frac{q^2}{p\cdot q}p_\mu p_\nu + \{pq\}_{\mu\nu}
\right](a^{(i)}_4 + b^{(i)}_4\lslash{v})\nonumber\\* 
&&\mbox{} + \left[
-\frac{q\cdot v}{p\cdot q}p_\mu p_\nu -\frac{p\cdot q}{q\cdot v}v_\mu v_\nu
+ \{pv\}_{\mu\nu}
\right](a^{(i)}_5 + b^{(i)}_5\lslash{v})\nonumber\\*
&&\mbox{} + \left[
-q\cdot v\eta_{\mu\nu} - \frac{q^2}{q\cdot v}v_\mu v_\nu + \{qv\}_{\mu\nu}
\right](a^{(i)}_6 + b^{(i)}_6\lslash{v})\nonumber\\* 
&&\mbox{} + \left[
-\lslash{q}\eta_{\mu\nu} + \frac{q^2}{(p\cdot q)^2}\lslash{q}p_\mu p_\nu
- \frac{q^2}{p\cdot q}\{p\gamma\}_{\mu\nu} + \{q\gamma\}_{\mu\nu}
\right](a^{(i)}_8 + b^{(i)}_8\lslash{v})\nonumber\\*
&&\mbox{} + \left[
\frac{q\cdot v}{(p\cdot q)^2}\lslash{q}p_\mu p_\nu -
\frac{\lslash{q}}{q\cdot v}v_\mu v_\nu -
\frac{q\cdot v}{p\cdot q}\{p\gamma\}_{\mu\nu} + \{v\gamma\}_{\mu\nu}
\right](a^{(i)}_9 + b^{(i)}_9\lslash{v})\,.
\end{eqnarray}
Some of the terms in \Eq{Gammagravmedium} vanish for an on-shell
graviton. However, the terms proportional to
the coefficients $a^{(i)}_{5,9}$ and $b^{(i)}_{5,9}$ do not
vanish and therefore the physical amplitude can be non-zero
in a medium.

\section{Conclusions}
We have considered various decay processes which are known
to be forbidden in the vacuum by a combination of requirements
such as angular momentum conservation,
Lorentz invariance or gauge invariance. As we showed, such processes
can occur in a medium, such as a thermal background of particles,
despite the fact that the medium may be homogeneous and isotropic.
To be precise, we carried out a model-independent analysis of
the vertex function for such processes in terms of a set of form factors,
and showed that the amplitude can be non-zero while remaining consistent
with the symmetry principles mentioned above.
The results simulate Lorentz symmetry violating effects, although in this
case they arise from completely Lorentz-invariant physics.

The idea that the Lorentz symmetry is not exact, and
the possible physical consequences of this,
has been of interest and an active field of research in the recent times
\cite{Carroll:1989vb,Coleman:1998ti,Jackiw:1999yp,Fogli:1999fs,%
Lorentzviolreviews}. The model-independent parametrization performed
in the present work is useful in these contexts as well. Firstly,
it can help to discriminate between the effects produced
by genuine Lorentz invariance violation at a fundamental level,
from similar effects that may arise even if it is not really violated.
Secondly, the calculation of the form factors that we have defined
require the specification of the background for the physical
situation at hand, but otherwise does not depend on any new physics beyond
the standard model. Therefore, the results of such calculations can be
used as benchmark values with which to compare the results
of similar calculations in the context of models of genuine Lorentz
symmetry violation.


\end{document}